\documentclass{article}
\usepackage[numbers]{natbib}
\usepackage[a4paper,margin=1in]{geometry}

\usepackage{graphicx}
\usepackage{dcolumn}
\usepackage{bm}

\usepackage{float}
\usepackage{subcaption}
\usepackage[margin=1in]{geometry}
\usepackage[colorlinks=true, citecolor=blue, linkcolor=blue, urlcolor=blue]{hyperref}
\usepackage{svg}
\usepackage[utf8]{inputenc}
\usepackage[T1]{fontenc}
\usepackage{mathptmx}
\usepackage{etoolbox}
\usepackage{lineno}
\usepackage{hyperref}
\usepackage{amsmath}
\usepackage{booktabs}

\hypersetup{
    colorlinks=true,
    linkcolor=blue,
    filecolor=magenta,      
    urlcolor=blue,
    citecolor=blue,
    }

\makeatletter
\def\@email#1#2{%
 \endgroup
 \patchcmd{\titleblock@produce}
  {\frontmatter@RRAPformat}
  {\frontmatter@RRAPformat{\produce@RRAP{*#1\href{mailto:#2}{#2}}}\frontmatter@RRAPformat}
  {}{}
}%
\makeatother
\begin{document}

\begin{center}

\textbf{Aerodynamic Drag and Heat Transfer Corrections for Dehydrated Pollen Particles: CFD-Based Modeling of Airborne Allergen Transport in Smart Urban Environments}

\vspace{2ex}

\textbf{Omar Hamad}$^{1,2}$, \textbf{Samer Ali}$^3$, \textbf{Mahmoud Khaled}$^{4,5}$ and \textbf{Talib Dbouk}$^{1*}$

\vspace{2ex}

$^1${CNRS, CORIA, UMR 6614, University of Rouen Normandy, F-76000 Rouen, France}

$^2${Energy \& Thermal Fluid Group, The International University of Beirut ${(BIU)}$, Mount Lebanon, Lebanon}

$^3${Université de Lille, Institut Mines-Télécom, Université d'Artois, Junia, ULR 4515-LGCgE, Laboratoire de Génie Civil et géo-Environnement, F-59000 Lille, France}

$^4${Energy \& Thermal Fluid Group, Lebanese International University (LIU), Bekaa, Lebanon\looseness=-1}

$^5${College of Engineering \& Architecture , Gulf University for Science and Technology, Kuwait\looseness=-1}

\vspace{2ex}

$^*$\textbf{Corresponding author}: \href{mailto:talib.dbouk@coria.fr}{talib.dbouk@coria.fr}
\end{center}

\begin{abstract}
\noindent\textbf{\large Abstract}\\[0.5em]
Airborne pollen transport is a key concern for urban air-quality assessment, allergy-risk forecasting, and smart-city planning. However, conventional dispersion models generally assume smooth spherical particles, neglecting how pollen dehydration alters particle morphology and impacts aerodynamic and thermal behavior. To address this gap, this study presents, for the first time, advanced CFD simulations evaluating the aerodynamic drag forces and convective heat transfer of realistically dehydrated (dry) pollen particles. Investigations are conducted at Reynolds numbers ($0.1 \leq \mathrm{Re_p} \leq 15$) at the particle's scale corresponding to realistic atmospheric wind speeds ranging from 0.27 to 30 km/h. The findings reveal that dry pollen particles exhibit drag coefficients 8\% to 15\% higher than those predicted for hydrated pollen spherical particles. Conversely, their Nusselt numbers are 5\% to 15\% lower than those for hydrated pollen particles. These considerable deviations confirm that conventional spherical correlations are inadequate for simulating dry pollen Lagrangian transport and evaporation. These findings highlight the need to account for realistic dehydrated shapes when modeling airborne allergen transport in urban environments.
\end{abstract}


\section{\label{sec:intro}Introduction}
Flow and plants is a very emerging field of research as described recently in a special issue in Physics of Fluids by Dbouk and Drikakis 2024 \cite{Dbouk2024}.

In plant biology, airborne pollen grains are small particles released from the male part of a flower that fertilize the female ovule, but these represent one of the most significant biological aerosols affecting human health worldwide. Birch pollen (\textit{Betula pendula}) is one of these airborne pollen grains that particularly notorious, being the second most common cause of tree pollen allergy after oaks, affecting millions of people across Europe, North America, and Asia. With climate change prolonging pollen seasons and increasing atmospheric pollen concentrations, the need for accurate predictive models of pollen dispersion and transport has never been more urgent.

Understanding the aerodynamic behavior of non-spherical and highly irregular biological particles is crucial for predicting accurately their transport, settling, drying process and dispersion in the atmosphere. Pollen grains, which of 20~$\mu$m mean diameter, feature highly irregular surface morphologies that can deviate significantly from the ideal spherical geometries assumed in most classical drag and heat transfer correlations \cite{schiller1933, Ranz1952, clift1978}. These geometric deviations, e.g. due to dehydration of pollen, can introduce substantial errors when predicting pollen dispersion and deposition rates, directly undermining the accuracy of health risk assessments in urban environments \cite{Dbouk2021, Dbouk2022, Dbouk:2026}.

For smooth spherical particles, the drag coefficient in the low-Reynolds-number regime is accurately described by Stokes' law \cite{stokes1851} for $\mathrm{Re_p} \ll 1$. For moderate Reynolds numbers ($0.1 \le \mathrm{Re_p} \le 400$), the Schiller-Naumann correlation \cite{schiller1933}, which serves as the standard drag model in conventional Lagrangian particle-tracking frameworks, provides an excellent approximation, while Clift et al. \cite{clift1978} established robust piece-wise formulations across a wider range. Similarly, heat transfer from spherical particles is traditionally modeled using the Ranz-Marshall correlation \cite{Ranz1952} for $\mathrm{Re_p} \le 200$, a standard formulation widely implemented in multiphase Computational Fluid Dynamics (CFD) solvers. However, the validity and direct applicability of these well-established spherical correlations to highly irregular, dehydrated biological particles like birch pollen remains an open scientific question.

Recent studies highlighted the critical role of high-fidelity transport modeling in public health frameworks. For instance, Dbouk and Drikakis \cite{Dbouk2021} demonstrated that airborne pollen grains can act as physical vectors for transmitting viral particles, while Dbouk et al. \cite{Dbouk2022} utilized advanced transport modeling to map pollen allergy risks in urban environments. These findings emphasized that errors in predicting aerodynamic drag or Nusselt numbers translate directly into flawed dispersion patterns and compromised epidemiological risk metrics.

 While developing precise empirical correlations is essential, classical experimental methods such as wind tunnel testing \cite{Ranz1952} and Particle Image Velocimetry (PIV) encounter severe limitations when applied to micro-scale particles of such complex, variable morphology due to evaporation (dehydration). Consequently, Reynolds-Averaged Navier-Stokes (RANS) modeling in CFD for Knudsen number $Kn\ll 1$, the continuum flow regime allow CFD to be thus cost-effective alternative capable of resolving detailed local aerodynamic and thermal fields across diverse flow conditions.

\begin{figure}[H]
    \centering
    \includegraphics[width=0.5\textwidth]{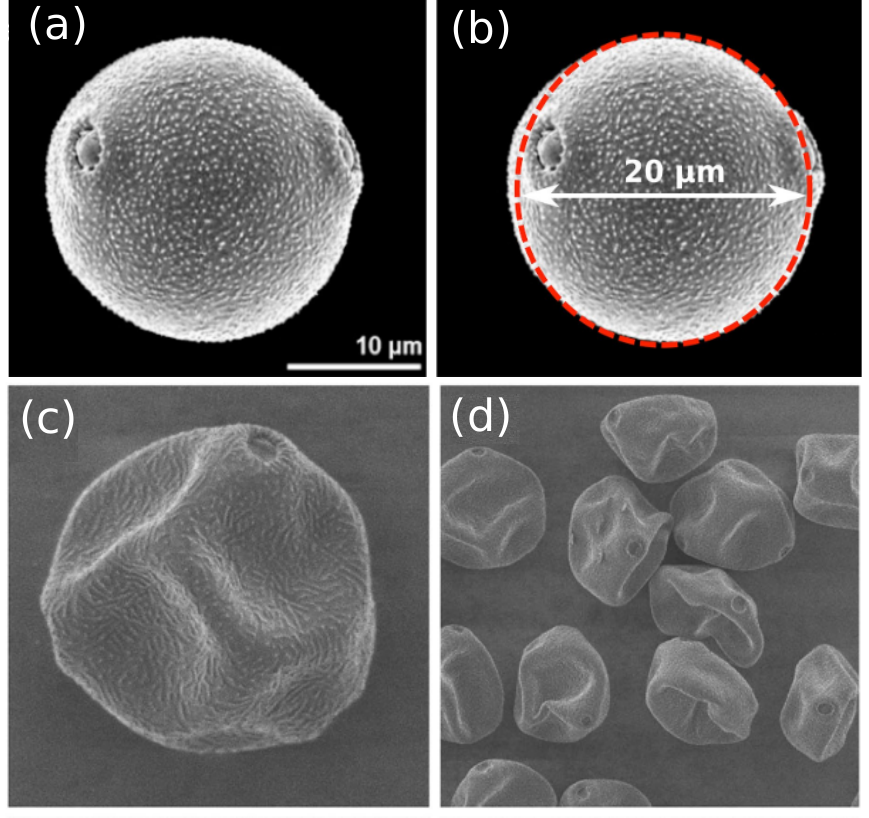}
    \caption{Morphological characterization of silver birch (\textit{Betula pendula}) pollen grain by scanning electron microscopy (SEM). (a) Hydrated pollen grain; (b) Approximated spherical shell of 20 $\mu$m diameter; (c) SEM of dry pollen grain; SEM image of multiple dry pollen grains. Author generated images from open-access CC BY \href{http://creativecommons.org/licenses/by/4.0/}{Creative Commons Attribution 4.0 International License}. (a) and (b) author generated images, see \cite{Dbouk2022, paldat_betula_2016}. (c) and (d) author generated images, see \cite{Depciuch2018BirchFTIR}.} 
    \label{fig:pollen_morphology}
\end{figure}

The present study investigates the applicability of classical spherical drag and Nusselt number correlations to realistic birch pollen morphology using high-fidelity CFD simulations. As shown in Fig.~\ref{fig:pollen_morphology}, which reproduces real scanned images of \textit{Betula} pollen grains from the PalDat database \cite{paldat_betula_2016}, the complex irregular features of the natural particle are faithfully preserved. The digital reconstruction represents the pollen grain in its dehydrated, dry state, capturing the collapsed outer shell characteristics while maintaining physical surface area constraints.
     
The remainder of this manuscript is organized as follows: Section~\ref{sec:Math} establishes the underlying CFD mathematical formulations and governing equations. Section~\ref{sec:level2} describes the high-fidelity numerical methodology and computational domain setup. Section~\ref{sec:level3} presents the computed drag coefficients, aerodynamic forces and Nusselt numbers alongside a comparative analysis against spherical correlations. Finally, Section~\ref{sec:level4} summarizes the key findings and outlines future research directions.

The originality of the present work lies in moving beyond the classical assumption of smooth spherical pollen particles by explicitly considering the realistic dehydrated morphology of dry pollen grains. Unlike previous studies that rely on standard spherical drag and heat-transfer correlations, this study evaluates, for the first time, morphology-resolved CFD-based aerodynamic forces and Nusselt numbers of dry airborne pollen particle under low-Reynolds-number atmospheric conditions. By quantifying the significant deviations from classical spherical laws, this work provides a new physical basis for improving pollen transport, deposition, and dispersion predictions in urban environments, with direct relevance to allergy-risk forecasting and smart-city air-quality assessment.

\section{\label{sec:Math}Numerical Methodology and Mathematical Formulations}

\subsection{\label{sec:Math0}Governing Aerodynamic Equations}
In the present work we will be working with heated forced convective flow (20–60 $^{\circ}$C). Knowing that at Mach < 0.3, the flow is essentially incompressible from a velocity standpoint, but
a compressible solver "rhoSimpleFoam" in OpenFOAM is applied due to temperature-induced density changes that are relatively non-negligible (up to 15\%).
The steady-state, compressible fluid flow fields surrounding non-spherical biological geometries are governed by the stationary, laminar Navier--Stokes equations in the continuum regime; e.g. Knudsen number $Kn \ll 1$, $O(10^{-3})$. 

The solved conservation equations of mass and momentum are defined as the following:

\begin{equation}
\nabla \cdot (\rho \mathbf{U}) = 0,
\end{equation}

\begin{equation}
\nabla \cdot (\rho \mathbf{U}\otimes\mathbf{U})
=
-\nabla p
+\nabla\cdot\boldsymbol{\tau}
\end{equation}

where $\rho$ represents the fluid density (a function of temperature), $\mathbf{U}$ is the velocity vector, $p$ is the static pressure, and $\boldsymbol{\tau}$ defines the molecular viscous stress tensor under a standard Newtonian fluid assumption. 

The viscous stress tensor $\boldsymbol{\tau}$ is defined as:

\begin{equation}
\boldsymbol{\tau}
=
\mu_{\mathrm{eff}}
\left[
\nabla\mathbf{U}
+
(\nabla\mathbf{U})^T
-
\frac{2}{3}
(\nabla\cdot\mathbf{U})\mathbf{I}
\right],
\end{equation}

where

\begin{equation}
\mu_{\mathrm{eff}}
=
\mu+\mu_t,
\end{equation}

with $\mu$ being the molecular viscosity and $\mu_t$ the turbulent eddy viscosity.

In the present work, because the micro-scale airflow around the pollen grain remains strictly within the low-Reynolds-number laminar regime ($\mathrm{Re_p} \le 15$), no turbulence closure model is applied ($\mu_t = 0$).

\subsection{\label{sec:Math1}Energy Conservation}
To accurately resolve the non-isothermal boundary layer variations arising from temperature differences between the particle and the fluid, the thermal energy field is solved concurrently. Reflecting the native implementation of the solver framework (\texttt{rhoSimpleFoam}) in OpenFOAM, the energy transport equation is formulated explicitly using sensible enthalpy ($h$), balancing steady-state convective advection of total enthalpy against thermal diffusion, while neglecting viscous dissipation and pressure work due to the low-speed flow regime:

\begin{equation}
\nabla\cdot(\rho h\mathbf{U})
=
\nabla\cdot
\left(
\alpha_{\mathrm{eff}}
\nabla h
\right)
+
\mathbf{U}\cdot\nabla p
\label{eq:energy}
\end{equation}

where $\alpha_{\mathrm{eff}}$ represents the effective thermal diffusivity. For this strictly laminar flow regime, it simplifies directly to molecular thermal diffusivity:

\begin{equation}
\alpha_{\mathrm{eff}} = \alpha_{\mathrm{laminar}} = \frac{\nu}{\mathrm{Pr}},
\end{equation}

incorporating the molecular kinematic viscosity of air ($\nu = 1.51 \times 10^{-5}~\mathrm{m^2/s}$), where $\mathrm{Pr}$ represents the molecular Prandtl number, initialized as $\mathrm{Pr} = 0.71$ to reflect standard ambient air conditions.

\subsection{\label{sec:Math2}Theoretical Boundary Layer Baselines}
To establish baseline numerical accuracy before evaluating complex, irregular pollen geometries, the computed transport coefficients on a smooth sphere are systematically benchmarked against canonical analytical and empirical correlations from the literature. 

Hydrodynamic drag is quantified non-dimensionally via the drag coefficient ($\mathrm{C_d}$). Under creeping flow conditions ($\mathrm{Re}_p \ll 1$), the fundamental analytical baseline is dictated by Stokes' drag law~\cite{stokes1851}:

\begin{equation}
\mathrm{C_d} = \frac{24}{\mathrm{Re_p}}.
\label{eq:stokes}
\end{equation}

To account for finite inertia effects at intermediate Reynolds numbers ($0.1 < \mathrm{Re}_p < 400$), the empirical correlation of Schiller and Naumann~\cite{schiller1933} is considered:

\begin{equation}
\mathrm{C_d} = \frac{24}{\mathrm{Re_p}} \left( 1 + 0.15 {\mathrm{Re_p}}^{0.687} \right).
\label{eq:schiller_naumann}
\end{equation}

For low to moderate Reynolds numbers at the particle's scale, the piecewise drag correlation of Clift et al ~\cite{clift1978} is employed which is based on a comprehensive review of experimental data and extends the validity down to creeping flow conditions including :

\begin{equation}
\label{eq:clift_piecewise}
C_d=
\begin{cases}
\dfrac{24}{\mathrm{Re}_p}
\left[1+0.1315\,\mathrm{Re}_p^{0.82-0.05\log_{10}(\mathrm{Re}_p)}\right],
& 0.01 \le \mathrm{Re}_p \le 20,\\[1ex]
\dfrac{24}{\mathrm{Re}_p}
\left(1+0.1935\,\mathrm{Re}_p^{0.6305}\right),
& 20 < \mathrm{Re}_p \le 260.
\end{cases}
\end{equation}

Concurrently, local convective heat transfer across solid-to-fluid interfaces is bench-marked against the foundational Ranz--Marshall Nusselt Number $(Nu)$ correlation~\cite{Ranz1952}:

\begin{equation}
\text{Nu} = 2 + 0.6 \mathrm{Re_p}^{1/2} \mathrm{Pr}^{1/3},
\label{eq:ranz_marshall}
\end{equation}

\subsection{\label{sec:Math3}Non-Dimensional Characterization Metrics}
Following this canonical baseline framework, the steady streamwise drag force component ($F_x$, aligned with the free-stream velocity vector $\mathbf{U}_\infty$) and the total integrated net surface heat flux ($\bar{Q}_w$) resolved across the mesh boundaries of the reconstructed pollen geometries are non-dimensionalized. The extracted drag coefficient ($\mathrm{C_d}$) and average Nusselt number ($\mathrm{Nu}$) are evaluated using the surface-area-equivalent diameter ($d_{\mathrm{eq}}$) as the unique characteristic length scale:

\begin{equation}
\mathrm{C_d} = \frac{8 |\mathbf{F}_x|}{\rho_{\infty} |\mathbf{U}_{\infty}|^2 \pi  d_{\mathrm{eq}}^2},
\label{eq:cd_calc}
\end{equation}

\begin{equation}
\mathrm{Nu} = \frac{h\, d_{\mathrm{eq}}}{\mathrm{k}_{\infty}},
\label{eq:nu_calc}
\end{equation}

where $h = \bar{Q}_w / (A_s \Delta T)$ represents the area-averaged convective heat transfer coefficient across surface area $A_s$. Free-stream fluid density, approach velocity, and thermal conductivity are denoted by $\rho_{\infty}$, $\mathbf{U}_{\infty}$, and $k_{\infty}$, respectively.

\section{\label{sec:level2}COMPUTATIONAL METHOD}
Over the past few decades, CFD has grown into a core method for analyzing fluid behavior, offering a reliable alternative to costly experiments specially of large scale\cite{Wendt2009}. Instead of relying on continuous differential equation, a conversion of the governing Navier-Stokes equations into discretized algebraic forms to simulate complex flows across different scales \cite{Versteeg2007}. At this heart, this approach relies on solving the fundamental of fluid motion. As outlined by Anderson in the classic introductory tex, these equations mathematically track three strict physics laws : mass conservation, Newton;s second law, and energy conservation \cite{Wendt2009}. Developing and using these methods practically would be impossible without high speed digital computers. Because these Simulations require processing millions of repetitive calculations, manual computations is out of the question. This means our ability to solve highly detailed fluid problems grows side by side with computing hardware, especially when it comes to memory storage and CPU processing speeds. It is exactly why the CFD community is still one of the biggest drivers behind modern supercomputer development. An open-source \textbf{OpenFOAM (Open Field Operation and Manipulation)} platform has been selected to handle these computations \cite{Weller1998}.

\subsection{Overview of OpenFOAM}

All numerical simulations in this study were performed using the open-source CFD software \textbf{OpenFOAM} \cite{Weller1998}. OpenFOAM implements the cell-centered \textbf{Finite Volume Method (FVM)} \cite{Versteeg2007} to discretize and solve the steady, compressible, non-isothermal Navier--Stokes equations via the solver \texttt{rhoSimpleFoam}.

The three-dimensional (3D) unstructured computational meshes enclosing both the baseline smooth spheres and the reconstructed non-spherical pollen architectures were generated utilizing OpenFOAM's \texttt{snappyHexMesh} utility. Grid refinement was systematically controlled near the particle shell to ensure rigorous boundary layer resolution..

\subsection{Regular Shaped objects: Spherical Particle – CFD Solver Validation}

To validate the accuracy of the numerical solver and mesh configuration prior to simulating complex pollen geometries, a benchmark CFD case for flow around a smooth sphere was executed across the low-Reynolds-number regime ($\mathrm{Re}_p \le 15$). The computed drag coefficients ($\mathrm{C_d}$) and Nusselt numbers ($\mathrm{Nu}$) were compared directly against the baseline analytical and empirical correlations established in Section ~\ref{sec:Math2} ~\ref{eq:clift_piecewise} \& ~\ref{eq:ranz_marshall} , respectively.

Table \ref{tab:table1} \& \ref{tab:table2} show some quantitative validation results for a fluid flow around a sphere compared to results reported in the literature.

\begin{table}[H]
\caption{\label{tab:table1}
Quantitative validation of the present CFD solver configuration applied to an air flow around a smooth sphere. Drag coefficient ($\mathrm{C_d}$) predictions at low Reynolds numbers ($\mathrm{Re_p}$) compared to results reported in the literature by Clift et al. (1978)\cite{clift1978} and Suri and Patel \cite{Suri2024}. The Reynolds numbers values correspond to values of wind speed that frequently occur in an urban environment (e.g. 2.7 km/h most frequent to 135 km/h Storm force) to transport tiny particles some micrometers in size.}

\centering

\begin{tabular}{ccccc}
\hline
$\mathrm{Re_p}$ & \multicolumn{2}{c}{Clift et al. (1978)\cite{clift1978}} & \multicolumn{2}{c}{Suri \& Patel (2024)\cite{Suri2024}} \\
\cline{2-5}
   & $\mathrm{C_d}$ & Error (\%) & $\mathrm{C_d}$ & Error (\%) \\
\hline
1 &  27.16 & +$4.86$ & 27.20 & +$4.7$\\
5   & 7.03  & +$3.10$  & 7.10  & +$2.07$ \\
10  & 4.26  & +$1.88$  & 4.28  & +$1.39$ \\
50  & 1.57  & +$2.87$  & 1.57  & +$2.81$ \\
\hline
\end{tabular}

\end{table}

\begin{table}[H]
\caption{\label{tab:table2}
Quantitative validation of the present CFD solver configuration applied to an air flow around a smooth sphere. Comparison of the Nusselt number ($\mathrm{Nu}$) predictions for a smooth sphere at low Reynolds numbers. Values from Ranz and Marshall (1952)\cite{Ranz1952} and Suri and Patel (2024)\cite{Suri2024} are compared against the present CFD results. Percentage errors relative to the present CFD are shown below.}

\centering
\begin{tabular}{ccccc}
\hline
Re & \multicolumn{2}{c}{Ranz \& Marshall (1952)\cite{Ranz1952}} & \multicolumn{2}{c}{Suri \& Patel (2024)\cite{Suri2024}} \\
\cline{2-5}
   & $\mathrm{Nu}$ & Error (\%) & $\mathrm{Nu}$ & Error (\%) \\
\hline
    1   & 2.54 & $-2.35$ & 2.25 & $+10.2$ \\
    5   & 3.20 & $-6.88$ & 2.85 & $+4.36$ \\
    10  & 3.70 & $-6.49$ & 3.34 & $+3.52$ \\
    50  & 5.81 & $-3.27$ & 5.44 & $+3.19$ \\
\hline
\end{tabular}

\end{table}

\subsection{Irregular Shaped Dry Pollen Particle}
After validating the CFD solver against the spherical benchmark case in tables \ref{tab:table1} and \ref{tab:table2}, the present study is extended to irregular shaped non-spherical dry pollen particle. We adopted a real pollen particle and reduced its volume while preserving its original surface area. The three-dimensional morphology of silver birch (\textit{Betula pendula}) pollen was modeled based on high-resolution images from the PalDat database~\cite{paldat_betula_2016}. Birch pollen is a very common cause of tree pollen allergy (hay fever), second only to oaks, and is characterized by an irregular, tri-porate geometry. The baseline hydrated model was then deformed to represent dehydrated (dry) pollen, a process that reduces the volume while largely preserving the surface area, leading to morphological changes including partial infolding of the outer wall.

\begin{figure}[H] 
    \centering
    \includegraphics[width=\linewidth]{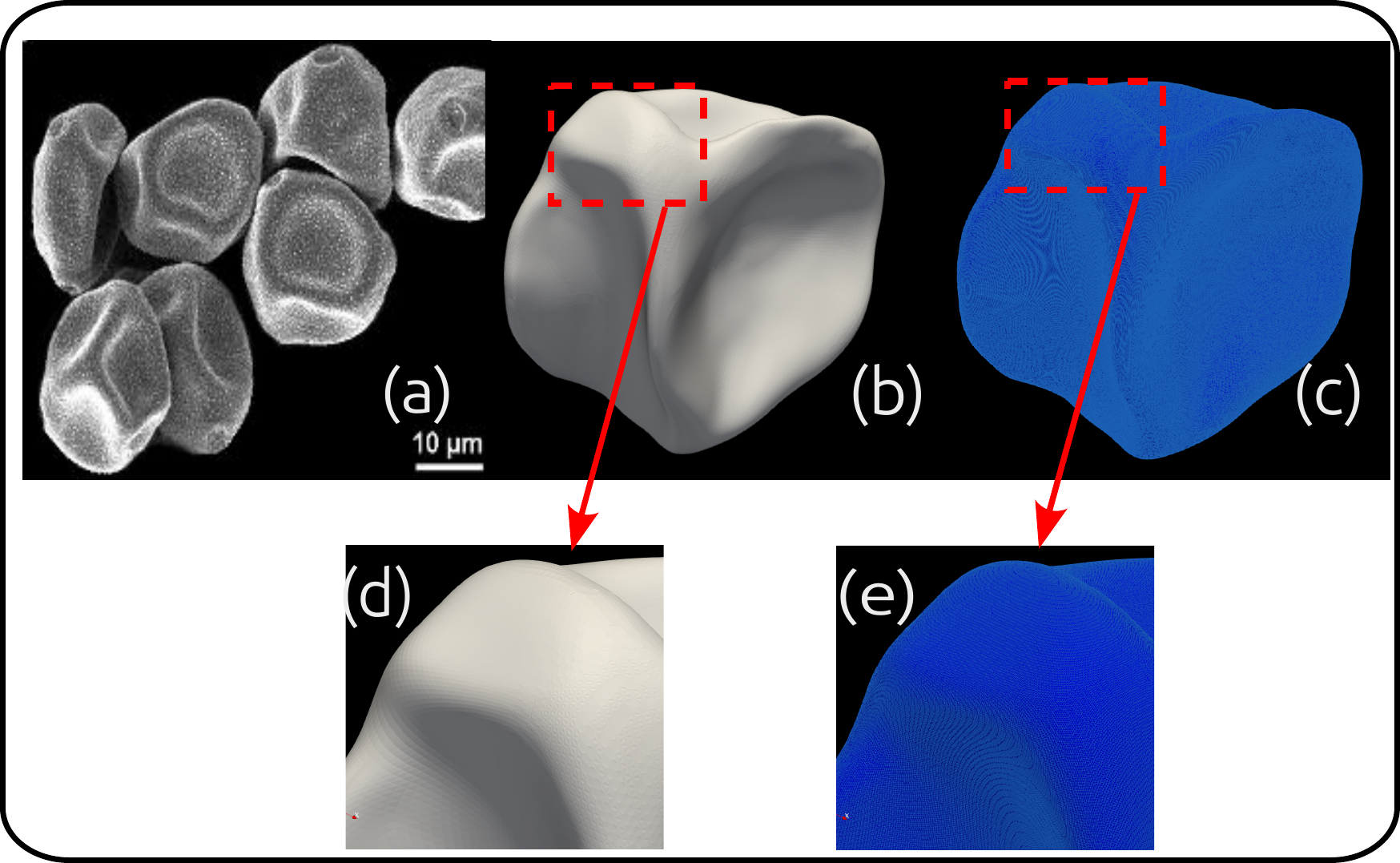} 
    \caption{3D morphological reconstruction of silver birch (\textit{Betula pendula}) pollen: (a) reference image from the SEM PalDat database~\cite{paldat_betula_2016}, (b) 3D STL (Stereolithography: Standard Tessellation Language) computer reconstruction of dry pollen, (c) dehydrated (dry) pollen Mesh generated in OpenFOAM, and (d, e) surface details of the STL and CFD Mesh used for computations, respectively.}
    \label{fig:Pollen_geometry}
\end{figure}
The 3D reconstruction of the dry birch pollen particle, as can be seen from figure \ref{fig:Pollen_geometry}, is made with high-quality. The computer aided design (CAD) reconstruction of this irregular pollen particle for CFD computations was very challenging, particularly to achieve enhanced surface quality suitable for accurate CFD numerical simulations. It is worth-noting that the surface area obtained is with almost less than 0.01 \% relative error that reflect a high percentage of identically obtained geometry and mesh.

\subsection{CFD Computational Domain and Mesh Resolution}
\subsection*{Boundary Conditions}
A free-stream boundary condition was applied on all outer faces of the computational domain, and the sphere surface was modeled as a no-slip wall thanks to a low Knudsen number Kn, $O(10^{-3})$. This setup was used to simulate flow around the sphere while minimizing boundary interference.
The adopted computational domain was defined with streamwise , transverse, and vertical dimensions satisfying $L = 1.5W = 1.5H = 24.3 d_p$, where $d_p$ denotes the particle equivalent diameter calculated from the surface area $A_s$.
At the internal fluid--solid boundary representing the particle shell, a no-slip condition ($\mathbf{U} = \mathbf{0}$) was imposed. Thermally, the surface of the particle was maintained at a constant elevated temperature of $T_p = 60^\circ\mathrm{C}$, while the incoming free-stream air was set to ambient conditions at $T_{\infty} = 20^\circ\mathrm{C}$ ($\Delta T = 40^\circ\mathrm{C}$) with a uniform approach velocity $\mathbf{U}_{\infty}$.
\subsection*{Mesh Generation}

Mesh generation was made by using the OpenFOAM modules \texttt{blockMesh} and \texttt{snappyHexMesh}, with local mesh refinement close the wall boundaries. Mesh sensitivity studies were made to ensure that the numerical solution is mesh-size-independent.

For these cases, the mesh was generated using two OpenFOAM preprocessing tools in sequence. First, \texttt{blockMesh} was employed to construct the background computational domain as a structured hexahedral grid with a Cartesian cell arrangement. Subsequently, \texttt{snappyHexMesh} was used to generate a body-fitted mesh around the pollen particle based on the STL geometry file located in the \texttt{constant/triSurface/} directory. The mesh refinement procedure consisted of castellated mesh generation followed by surface snapping to accurately capture the particle geometry.

A grid refinement study was conducted to ensure mesh-independent results. Three meshes with increasing cell counts were evaluated following the \textbf{Grid Convergence Index (GCI)} method based on Richardson extrapolation\cite{Celik2008}. 
Three unstructured meshes---coarse ($M_3$), medium ($M_2$), and fine ($M_1$)---were evaluated. To optimize computational resources while maintaining localized resolution, non-uniform grid refinement was utilized, yielding a medium-to-fine refinement ratio of $r_{21} = h_2/h_1 = 1.32$ and a coarse-to-medium refinement ratio of $r_{32} = h_3/h_2 = 1.27$.

Two primary engineering parameters representing the coupled aerodynamic and thermal fields were selected as target criteria for the Grid Convergence Index (GCI): the total stream-wise drag force ($F_x$) acting on the solid boundaries (comprising integrated pressure and viscous shear forces), and the surface-area-averaged heat transfer coefficient ($\bar{h}$) evaluated across the thermal exchange zone. 

By applying the transcendental iterative framework to account for the non-uniform refinement ratios, the apparent order of numerical accuracy ($p$) was determined to be [$1.37$] for the momentum field ($F_x$) and 0.638 for the thermal field ($\bar{h}$). The calculated fine-grid convergence index ($\text{GCI}_{21}$) yielded a numerical uncertainty of 2.7\% for the drag force and 2.6\% for the heat transfer whereas the calculated medium-grid convergence index  coefficient$\text{GCI}_{32}$) yielded a numerical uncertainty of 3.8\% for the drag force and 3.6\%. All values fall safely below the conservative 5\% threshold recommended for engineering verification. Furthermore, the convergence factors $\text{GCI}_{32} / (r_{21}^p \cdot \text{GCI}_{21})$ approached unity ($\approx 1.0$), confirming that the solutions strictly occupy the asymptotic range of convergence. Consequently, the coarse mesh ($M_3$) was selected for all production simulations. The grid parameters and convergence metrics are summarized in Table~\ref{tab:gci_aerodynamic}.


\begin{table}
\caption{\label{tab:gci_aerodynamic}Grid convergence parameters and spatial discretization error estimations for the aerodynamic field (drag force $f_x$).}
\centering
\begin{tabular}{lccc}
Parameter &  $M_3$ &  $M_2$  &  $M_1$  \\
\hline
$N$ & 6765980 & 13765637 & 31338640\\
$F_x$  & $9.37 \times 10^{-8}$ & $9.26 \times 10^{-8}$ & $9.17 \times 10^{-8}$ \\
\hline
     & \multicolumn{2}{c}{$M_3/M_2$} & $M_2/M_1$ \\ 
\hline
($r$) & \multicolumn{2}{c}{$r_{32} = 1.27$} & $r_{21} = 1.32$ \\ 

($e_a$) & \multicolumn{2}{c}{$e_{a,32} =$ 1.18\%} & $e_{a,21} =$ 0.98\% \\
 ($\text{GCI}$) & \multicolumn{2}{c}{$\text{GCI}_{32} =$ 3.8\%} & $\text{GCI}_{21} =$ 2.7\% \\ 
\hline
check & \multicolumn{3}{c}{$\text{GCI}_{32} / (r_{21}^p \cdot \text{GCI}_{21}) \approx 1.0$} \\
\hline
\end{tabular}
\end{table}

\begin{table}
\caption{\label{tab:gci_thermal}Grid convergence parameters and spatial discretization error estimations for the thermal field (average heat transfer coefficient $\bar{h}$).}
\centering
\begin{tabular}{cccc}
Parameter & $M_3$ (Coarse) & $M_2$ (Medium) & $M_1$ (Fine) \\
\hline
$N$ & 6765980 & 13765637 & 31338640\\
$\bar{h}$ ($\text{W/m}^2\text{K}$) & $3164$ & $3179$ & $3167$ \\
\hline
     & \multicolumn{2}{c}{$M_3/M_2$} & $M_2/M_1$ \\ 
\hline
($r$) & \multicolumn{2}{c}{$r_{32} = 1.27$} & $r_{21} = 1.32$ \\ 
$e_a$ (\%) & \multicolumn{2}{c}{$e_{a,32} = 0.47$ } & $e_{a,21} = 0.4$ \\
$\text{GCI}$ (\%) & \multicolumn{2}{c}{$\text{GCI}_{32} = 3.6$ } & \multicolumn{1}{l}{$\text{GCI}_{21} = 2.6$ } \\ 

\end{tabular}
\end{table}

The primary quantities of interest used to characterize the aerodynamic forces and heat transfer performance are the drag coefficient ($\mathrm{C_d}$) and the Nusselt number ($\mathrm{Nu}$). For the baseline numerical setup, a grid resolution consisting of approximately 6.8 million cells was utilized. To optimize computational efficiency while maintaining accuracy within the low Reynolds number regime ($Re \le 15$) investigated in this study, the strictly laminar boundary layers at the pollen surface were resolved directly by the localized grid refinement without the insertion of prism inflation layers. This mesh configuration ensures that the velocity and thermal gradients at the fluid-solid interface are captured accurately within the linear viscous regime.

The aerodynamic and heat transfer characteristics of the non-spherical biological particles are quantified utilizing key dimensionless parameters. The Reynolds number ($\mathrm{Re_p}$) at the particle's scale, which evaluates the ratio of inertial forces to viscous forces within the fluid domain, is defined as:

\begin{equation}
\mathrm{Re_p} = \frac{\rho_{\infty} U_{\infty} \mathrm{L_c}}{\mu},
\label{eq:Re}
\end{equation}

where $\rho_{\infty}$ is the fluid density, $U_{\infty}$ is the free-stream approach velocity, and $\mu$ is the dynamic viscosity of the fluid. The characteristic length ($\mathrm{L_c}$) is defined as the surface area-equivalent diameter ($d_{eq}$) of the irregular pollen particle geometry (dehydrated or dry pollen), calculated as follows:

\begin{equation}
\mathrm{L_c} = d_{eq} = \sqrt{\frac{A_s}{\pi}},
\label{eq:deq}
\end{equation}

where $A_s$ represents the total surface area of the pollen particle.

The Nusselt number, characterizing the convective heat transfer, is given by:
\begin{equation}
\mathrm{Nu} = \frac{h \mathrm{L_c}}{k},
\label{eq:Nu}
\end{equation}
where $h$ is the convective heat transfer coefficient and $k$ is the thermal conductivity of the fluid.\\

Finally, the drag coefficient is computed as:
\begin{equation}
\mathrm{C_d} = \frac{|F_x|}{\frac{1}{2} \rho_{\infty} U_{\infty}^2 A_{ref}},
\label{eq:Cd}
\end{equation}
where $F_x$ is the drag force integrated from pressure and viscous stresses on the particle surface, $U_{\infty}$ is the reference velocity, and $A_{ref}$ is the projected area of the particle that given by $A_{ref} = \frac{\pi d_{eq}^2}{4}$.

$\mathrm{L_c}$ is the characteristic length (equivalent diameter $d_{eq}$ from the particle surface area), $F_x$ is the drag force integrated from pressure and viscous stresses on the particle surface, and $A_{ref}$ is the projected area of the spherical particle.

\section{\label{sec:level3}Results and Discussions}
\begin{figure}[H]
    \centering
    \includegraphics[width=\linewidth]{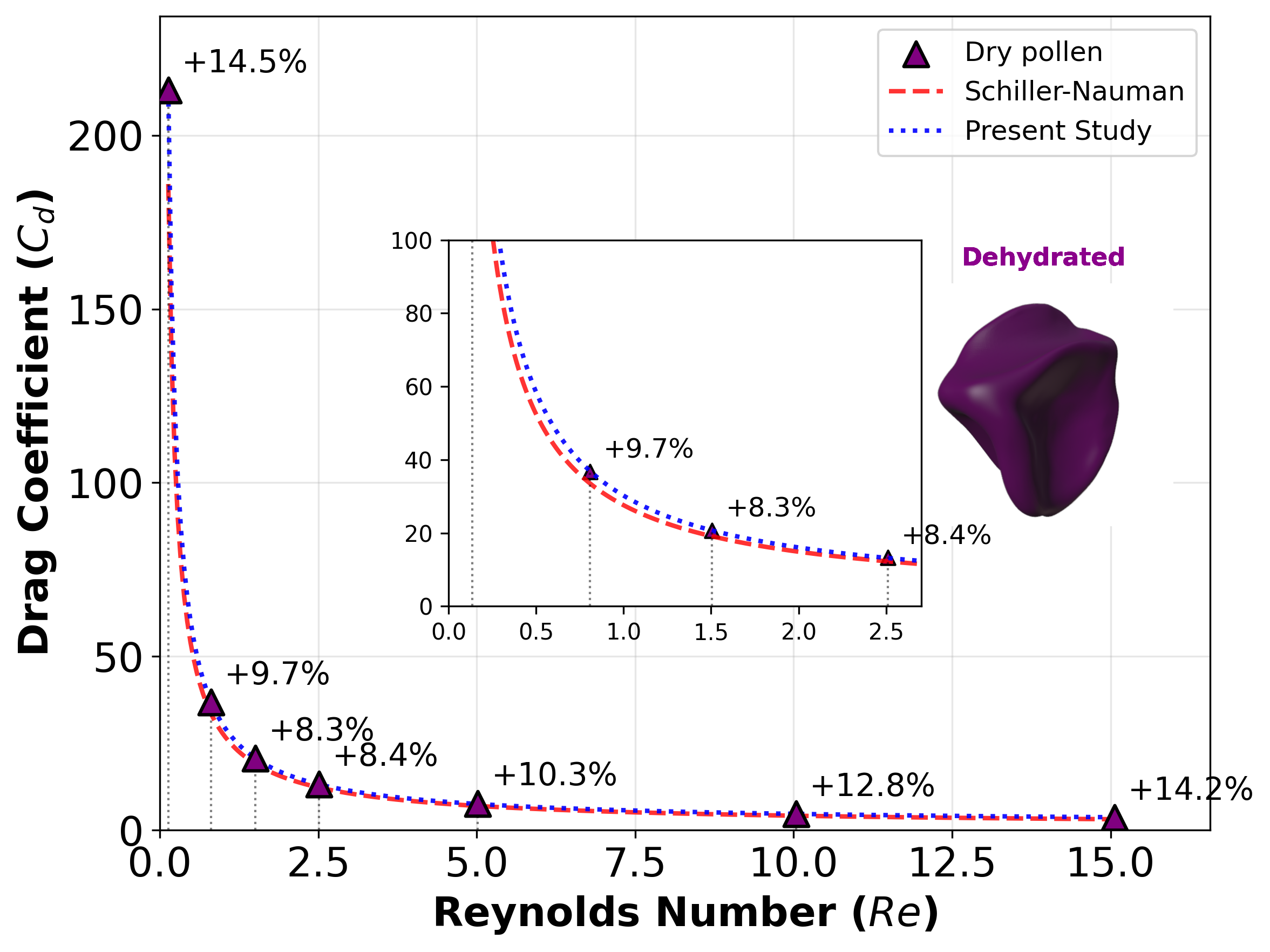}
    \caption{CFD simulation results for the drag coefficient ($\mathrm{C_d}$) of a dry (dehydrated) pollen irregular particle as function of Reynolds number ($\mathrm{Re}=\mathrm{Re_p}$). The results show percentages of deviation to the correlation established in literature bu Schiller and Naumann 1933 \cite{schiller1933}. The Reynolds number values correspond to wind speeds ranging between 0.27 and 30 km/h. }
    \label{fig:CD_pollen}
\end{figure}

The dry pollen particle shows a higher drag coefficient in Figure \ref{fig:CD_pollen} compared to the spherical pollen particle over the entire Reynolds-number range considered (0.1 $\leq \mathrm{Re_p} \leq$ 15). At the lowest Reynolds number, the deviation is already noticeable, and it remains positive for all cases, reaching values between  +8\% and +15\% depending on the flow condition. This means that the irregular dry pollen particle is subject to higher flow resistance than an equivalent sphere, even-though the characteristic diameter is the same.\\

\begin{figure}[H]
    \centering
    \includegraphics[width=\linewidth]{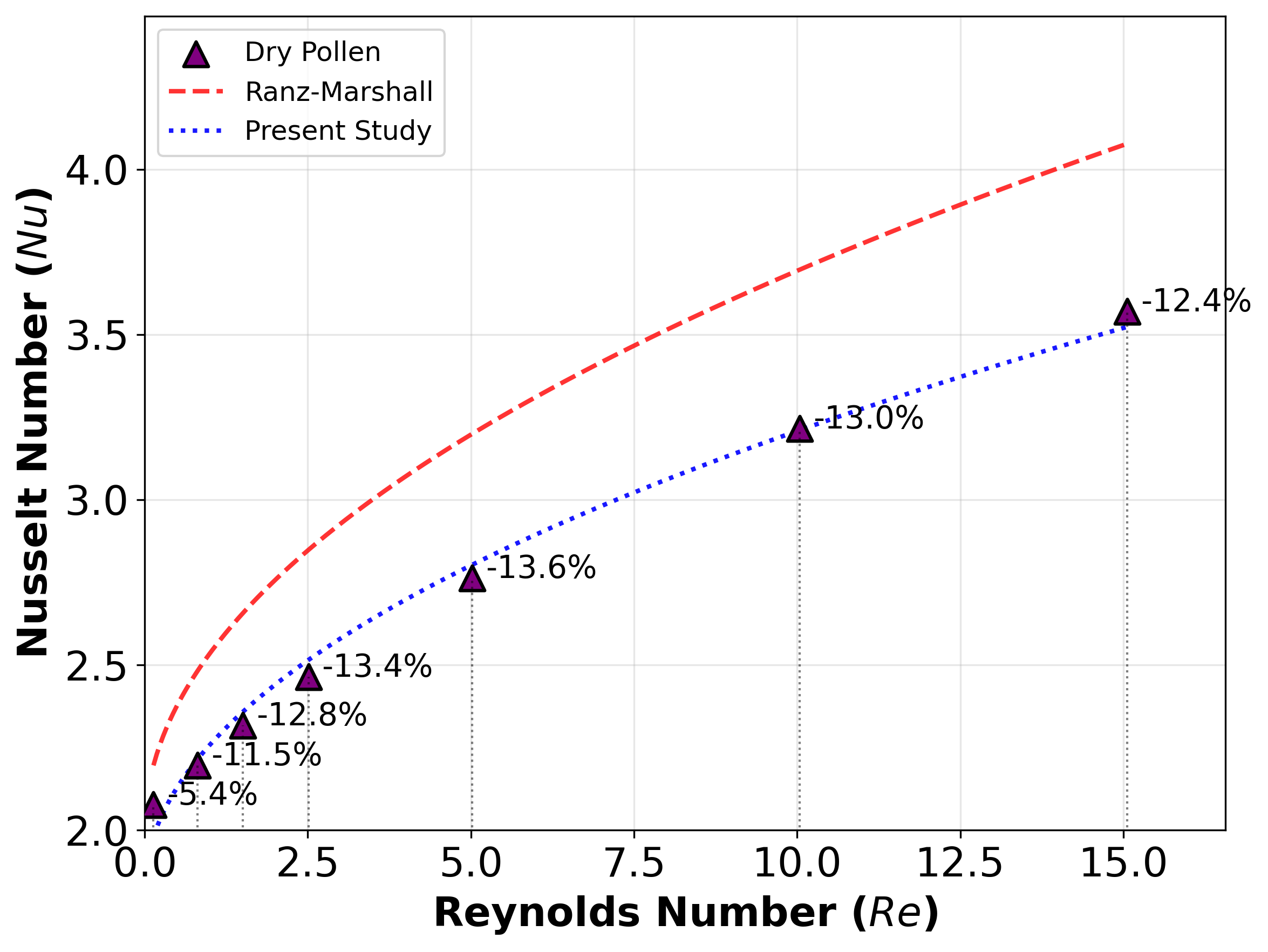}
    \caption{CFD simulation results for the Nusselt number of a dry pollen irregular particle as function of the Reynolds number ($\mathrm{Re}=\mathrm{Re_p}$) showing percentages of deviation to the correlation established in literature by Ranz and Marshall 1952 \cite{Ranz1952}. The Reynolds number values correspond to wind speeds ranging between 0.27 and 30 km/h.}
    \label{fig:Nu_pollen}
\end{figure}

The Nusselt number in Figure \ref{fig:Nu_pollen} follows the opposite trend. In every case, the dry pollen particle has a lower Nusselt number than the spherical hydrated pollen, with values between -5\% and -14\%. The difference is largest at high and intermediate Reynolds numbers, then it becomes slightly smaller at lower Reynolds numbers. This suggests that the irregular dry pollen geometry reduces the convective heat-transfer performance compared with the spherical reference. This is an extremely important conclusion that means that when pollen particles are under evaporation process, the evaporation mechanism decelerates with time and thus induces larger characteristic periods of evaporation compared to what is expected from/within spherical particles/droplets. This also means that using spherical correlations for both drag and evaporation for pollen grains can importantly underestimates the drag forces and significantly overestimates the evaporation strength.

\begin{figure}[H]
    \centering
    \includegraphics[width=\linewidth]{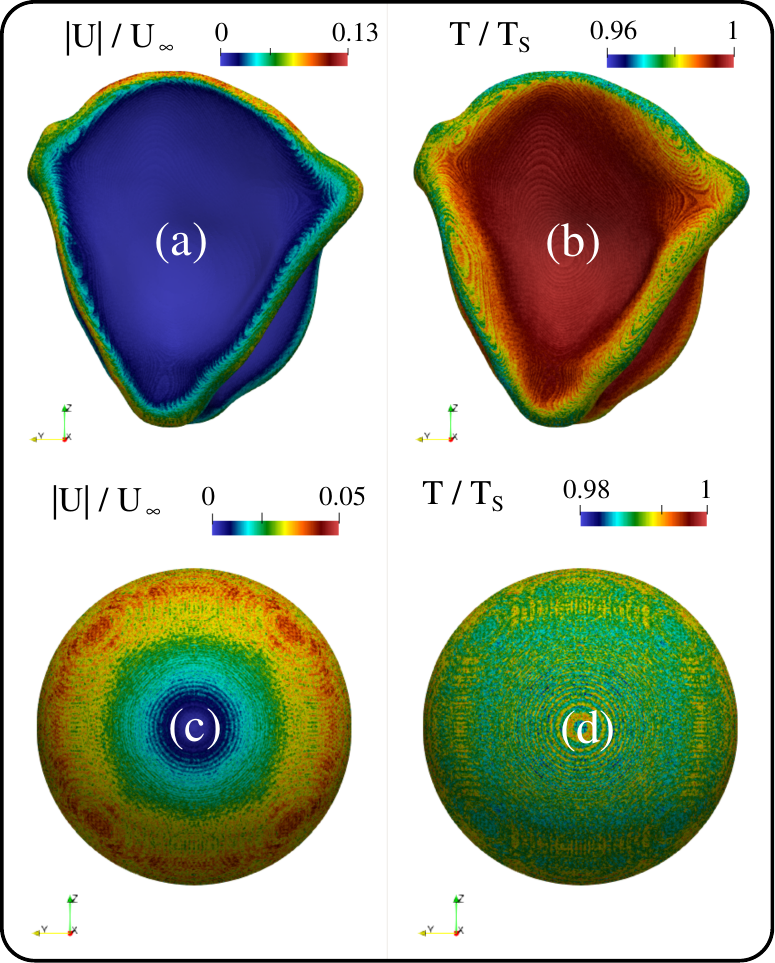}
    \caption{CFD simulation results at a near-surface fluid layer close to the pollen particle's surface. (a) dimensionless velocity ($U/U_{\infty}$) and (b) dimensionless temperature ($T/T_{S}$) for a dry irregular dehydrated pollen particle. (c) dimensionless velocity and (d) dimensionless temperature for a spherical hydrated pollen particle. The comparison provides a clear reference to quantify the hydrodynamic and thermal boundary layer alterations induced by biological surface irregularities of pollen particle.}
    \label{fig:layer_ratios}
\end{figure}

Figure~\ref{fig:layer_ratios} reveals a contrast between the uniform symmetric gradients velocity and temperature of a nearest layer of mesh of the baseline sphere and the highly distorted distributions surrounding the irregular dry pollen geometry. For the smooth pollen sphere (Figs.~\ref{fig:layer_ratios}c and \ref{fig:layer_ratios}d), both the velocity and thermal profiles exhibit concentric, predictable gradients, representing growth with symmetric localized thermal resistance. In contrast, the morphological depressions, ridges, and complex surface grooves of the biological dry pollen grain (Figs.~\ref{fig:layer_ratios}a and \ref{fig:layer_ratios}b) severely disrupt this canonical behavior. In the deep surface valleys of the pollen grain, the approaching fluid experiences localized stagnation, causing the velocity to drop significantly. This localized trapping of air creates a thick, stagnant thermal shielding layer within the hollows, severely restricting convective transport and increasing local thermal resistance. Consequently, the local heat flux drops dramatically within these valleys. Conversely, along the sharp protruding boundaries and structural ridges, the boundary layer is compressed by the oncoming free-stream flow, causing sharp temperature gradients and highly elevated local heat fluxes. However, because the low-flux stagnant depressions occupy a substantial portion of the total surface area of the reconstructed pollen geometry, their isolating effect outweighs the high-flux ridges. When integrated across the entire irregular surface, the net cumulative heat transfer is lower than that of a smooth sphere of equivalent volume, definitively accounting for the lower overall Nusselt numbers ($\mathrm{Nu}$) observed for the irregular dry pollen particle addressed in Figure \ref{fig:Nu_pollen}.

\begin{figure}[H]
    \centering
    \includegraphics[width=\linewidth]{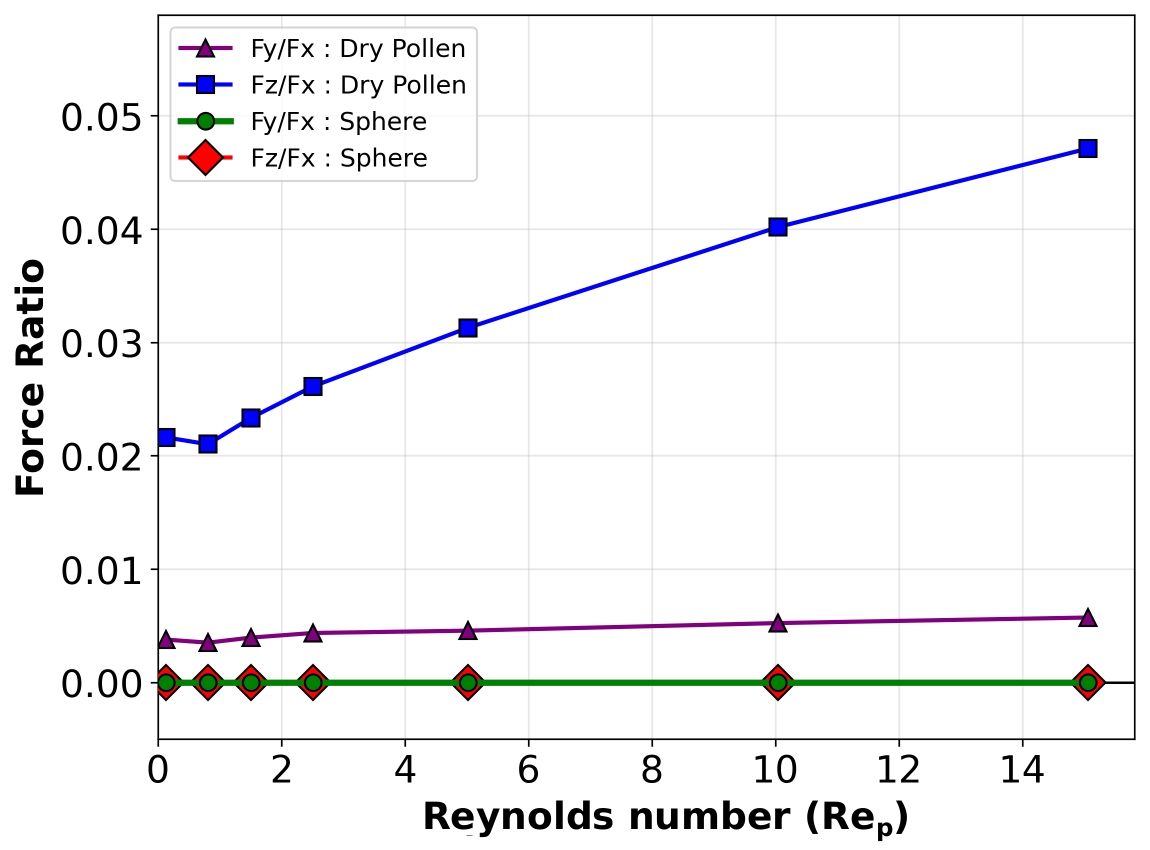}
    \caption{Comparison of the lateral-to-streamwise force ratios ($F_y/F_x$ and $F_z/F_x$) as a function of the Reynolds number ($\mathrm{Re}=\mathrm{Re_p}$) for a dry pollen grain and a baseline perfect sphere. Due to geometric symmetry, the sphere maintains zero lateral forces across the entire laminar regime, whereas the structural asymmetry of the dry pollen grain generates significant morphology-induced lift and side forces that scale with inertial effects.}    \label{fig:forces}
\end{figure}

To quantify the aerodynamic consequences of non-spherical irregular-shaped dry pollen particle, the dimensionless force ratios ($F_y/F_x$ and $F_z/F_x$) were plotted against the particle Reynolds number ($\mathrm{Re_p}$) ranging from $0.1$ to $15$ as shown in Figure \ref{fig:forces}. As a validation baseline, a smooth, perfectly symmetric sphere was simulated under identical flow conditions. As expected from analytical fluid mechanics, the sphere yields localized lateral and lift force ratios of exactly zero across the entire evaluated $\mathrm{Re_p}$ envelope, confirming that no asymmetric flow separation or lateral pressure gradients occur at low Reynolds numbers.

Conversely, the dry pollen particle exhibits distinct, non-zero force ratios, serving as a direct manifestation of its asymmetric surface features and complex morphology. While the side force ratio $F_y/F_x$ remains relatively low and stable (climbing minimally from $0.004$ to $0.006$), the lift-to-drag ratio $F_z/F_x$ displays a pronounced sensitivity to inertial effects. After a minor initial dip at $\mathrm{Re_p} = 1$, the $F_z/F_x$ ratio experiences a steady, linear growth, reaching approximately $0.047$ at $\mathrm{Re_p} = 15$. This demonstrates that at higher localized slip velocities, the lift force can account for nearly $5\%$ of the primary streamwise drag force. These findings highlight why conventional spherical approximations fail to capture true transport behaviors, as morphology-induced forces significantly alter particle lift and drift trajectories in low-Reynolds boundary layer flows.

\subsection{Importance of accurate transport correlations}

\begin{figure}[H]
    \centering
    \includegraphics[width=\linewidth]{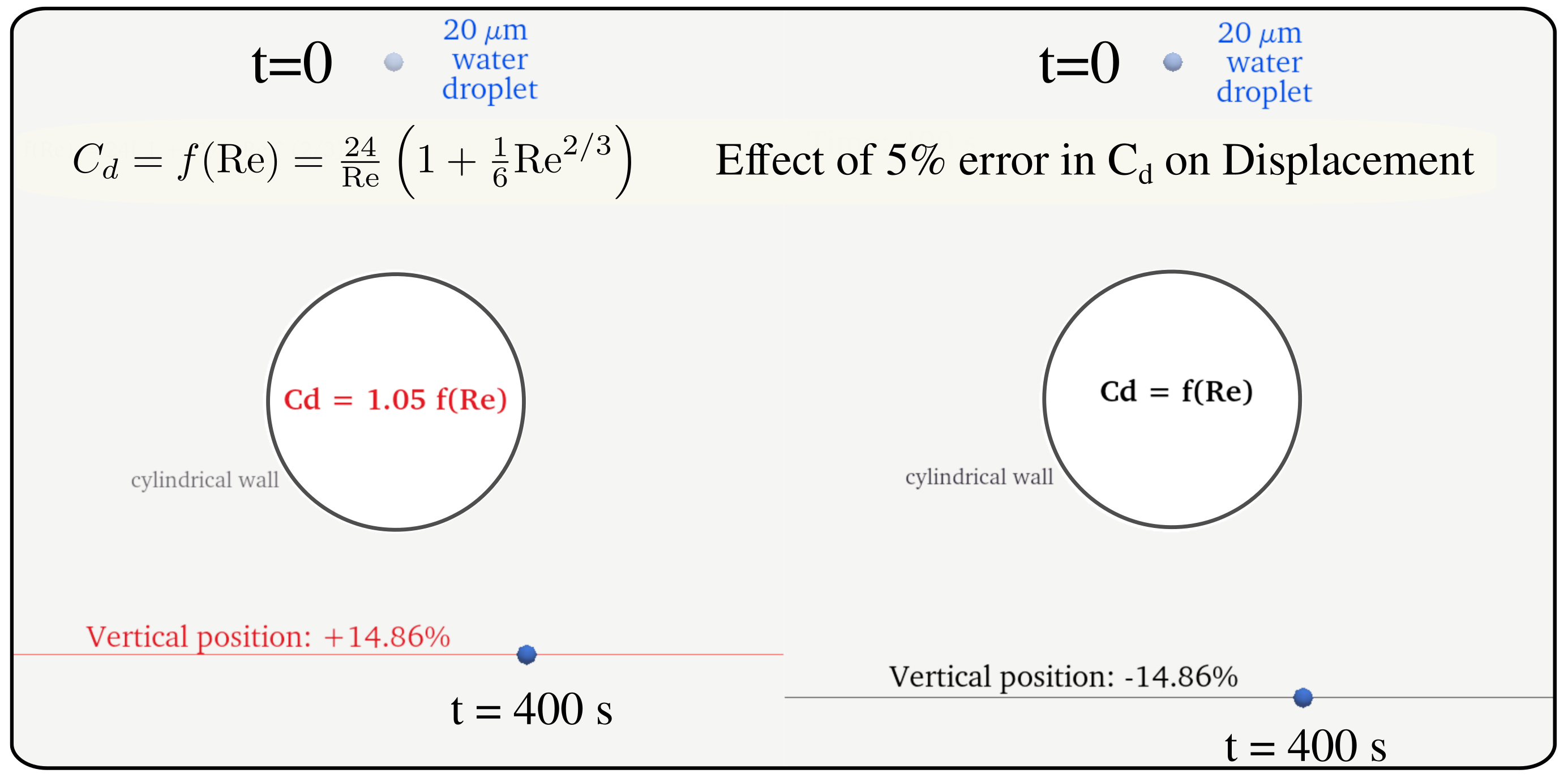}
    \caption{This figure illustrates the position of a 20 micrometer water droplet after 400 seconds employing Euler-Lagrange CFD with parcel's method. It clearly shows how only a 5\% error imposed to the drag coefficient law can importantly affect the final position of the droplet.}
    \label{fig:CD_error}
\end{figure}

Figure \ref{fig:CD_error} shows a good example about the importance of having accurate drag coefficient correlation to accurately predict the local dynamics of airborne pollen. In this example, it can be clearly seen that a simple +5\% deviation in the drag coefficient induces about 14.8\% error in the local displacement prediction of a settling water droplet (for example, a good mimic of an airborne moving spherical pollen particle) moving in still air between t=0 and t=400 seconds.

\section{\label{sec:level4}Conclusion and perspectives}

For the first time to our knowledge, it is shown how the dehydration of pollen (dry pollen) alter their morphology and thus the spherical drag correlations do not hold any more to predict the transport dynamics and evaporated mass of water content. Consequently, new aerodynamics forces and Nusselt numbers of dry pollen particle are developed and presented for the first time by employing advanced computational fluid dynamics (CFD) at different low Reynolds numbers ($0.1 \leq \mathrm{Re_p} \leq 15$) at the particle's scale that correspond to realistic atmospheric wind speeds ranging from 0.27 to 30 km/h. 
The results showed that an airborne dry pollen particle in an air flow does not behave exactly like a spherical one, even when the same characteristic diameter is employed. The shape irregularity increases the aerodynamic drag with non-zero force ratios, and reduces the Nusselt number which explains a deceleration in the evaporation process with time. This indicates that sphere-based correlations cannot fully capture the coupled momentum and thermal behavior of real airborne pollen particles undergoing dehydration or evaporation process. In another research paper, a focus will be given to developing new correlations that correspond to different airborne pollen shapes undergoing a dehydration process under different wind speeds taking into account the angle of rotation of the airborne pollen.


\section*{Acknowledgments}
The authors gratefully acknowledge the financial support of the Normandy Region (\href{https://www.normandie.fr/}{Région Normandie}) and the ANR (l'Agence nationale de la recherche) for funding our present research activities under: project "\href{https://transitions.univ-rouen.fr/projet-transition/}{TRANSITION} ExcellencES France 2030, ANR-23-EXES-0013" (la transition socio-écologique et la recherche sur les "multirisques" liés au changement climatique); project "DISPERSE" (Airborne pollutant particles dispersion close to an emitting source in an urban environment: Modeling and Simulations); and project "PAPA". The authors express their gratitude to the \href{https://www.coria.fr/en:}{CORIA} laboratory and the \href{https://www.univ-rouen.fr}{University of Rouen Normandy} for their institutional support. 
The authors acknowledge the international co-financial support of "LIU" (Lebanese International University, Lebanon) and the "BIU" (The International University of Beirut, Lebanon).
The authors thank the support of "CESM" \href{https://multirisques.univ-rouen.fr/}{Centre d’Expertise Scientifique sur le Multirisques} at the University of Rouen Normandy.
The authors also acknowledge the high-performance computing resources provided by \href{https://www.criann.fr}{CRIANN} data-center under projects number 2023013 and number 2025011. Finally, special thanks are extended to the \textbf{3D Pollen Project} and the PalDat database \cite{paldat_betula_2016} for providing the 3D models that greatly assisted in the geometric reconstruction and computer generation of the STL files used in the CFD simulations.

\section*{Data Availability Statement}

The data that support the findings of this study are available from the corresponding author upon reasonable request.

\section*{Authors Contributions}
\textbf{O.H.} Conceptualization, Methodology, Software, Visualization, Data, Analysis, Original Draft. \textbf{S.A.} Methodology, Co-Supervision, Analysis, Review \& Edit. \textbf{M.K.} Methodology, Co-Supervision, Analysis, Review \& Edit. \textbf{T.D.} Conceptualization, Methodology, Software, Supervision, Analysis, Review \& Edit, Funding, Project Administration.

\bibliographystyle{unsrtnat}
\bibliography{references.bib}

\end{document}